\documentstyle[preprint,aps,eqsecnum,epsf,floats]{revtex}

\newcommand{\PSbox}[3]{\mbox{\rule{0in}{#3}\includegraphics{#1}\hspace{#2}}}
\def\MgM{M\~gM}

\def\qslash{\not{\hbox{\kern-2pt $q$}}}
\def\delslash{\not{\hbox{\kern-2pt $\partial$}}}

\def\beq{\begin{equation}}
\def\eeq{\end{equation}}
\def\eeq{\end{equation}}
\def\bea{\begin{eqnarray}}
\def\eea{\end{eqnarray}}
\def\bq{\begin{quote}}
\def\eq{\end{quote}}
\def\lessim{\mathrel{\mathpalette\vereq<}}

\def\ddt{\frac{d}{dt}}
\makeatletter
\def\wtil{\widetilde}
\def\vereq#1#2{\lower3pt\vbox{\baselineskip1.5pt \lineskip1.5pt
\ialign{$\m@th#1\hfill##\hfil$\crcr#2\crcr\sim\crcr}}}
\makeatother

\begin{document}
\preprint{\vbox{\hbox{SLAC-PUB-8442}
		\hbox{UCSD/PTH-00-10}
                \hbox{hep-ph/0004210}}}

\title{\vbox{\vskip 0.truecm} The Superpartner Spectrum of Gaugino Mediation}

\author{\vbox{\vskip 0.truecm} Martin Schmaltz $^x$ and Witold Skiba $^y$}
\address{
          \vbox{\vskip 0.truecm}
          $^x$SLAC, Stanford University, Stanford, CA 94309 \\
          {\tt schmaltz@slac.stanford.edu} \\
          \vbox{\vskip 0.truecm}
         $^y$Department of Physics, University of California at San Diego, 
          La Jolla,  CA 92093 \\
          {\tt skiba@einstein.ucsd.edu} }
\maketitle

\begin{abstract}%

We compute the superpartner masses in a class of models with
gaugino mediation (or no-scale) boundary conditions at a scale between
the GUT and Planck scales. These models are
compelling because they are simple, solve the supersymmetric flavor
and CP problems, satisfy all constraints from colliders and cosmology, and
predict the superpartner masses in terms of very few parameters.   
Our analysis includes the renormalization group
evolution of the soft-breaking terms above the GUT scale. We show that
the running above the GUT scale is largely model independent and find that
a phenomenologically viable spectrum is obtained.   

\end{abstract}

\newpage

\section{Introduction}

The soft-breaking terms in the Minimal Supersymmetric Standard Model (MSSM)
need to have a very special form for the model to be viable.
Generic mass matrices for the squarks
and sleptons lead to unacceptably rapid flavor-changing and
lepton-number-violating processes. Similarly, the tri-linear
soft-breaking breaking A-terms also require fine tuning
for the MSSM to agree with experiment. It is important to explore models
which naturally solve the flavor fine-tuning problem.

The rates for flavor-changing processes induced by the soft masses are
proportional to the ratios of the off-diagonal mass squares to the diagonal
ones. In order to make such ratios small one needs to minimize the off-diagonal
terms or increase the flavor-preserving masses. Examples of natural scenarios
are gauge mediation \cite{gauge1,gauge2} and anomaly mediation
\cite{anomalymediation}, which produce diagonal mass
matrices due to the universality of the gauge coupling, and effective
supersymmetry~\cite{CKN}, which postulates large soft masses for
the first two generations, for which the experimental constraints are most
stringent. The popular minimal supergravity~\cite{SUGRA} model does
not contain a solution to the flavor problem, instead
one simply assumes that the higher-dimensional operators which
produce the scalar masses are flavor preserving.

It has been noticed that in ``gaugino-dominated'' models the flavor problem
is less severe~\cite{DKS}. If at a high scale gaugino masses are larger than
other soft parameters at that scale, then at low energies the soft
masses consist of large flavor-conserving masses with smaller
flavor-violating components. 
This is because the renormalization group running induces
(positive!) universal soft scalar masses
proportional to the gaugino masses. The generated scalar masses are
flavor universal because couplings to gauginos are
generation-independent. The most appealing gaugino-dominated
scenarios have no soft masses for squarks and
sleptons and no A-terms at a high scale $M_{BC}$.
Since at the high scale there are no masses and no A-terms, the only sources
of flavor violation are the Yukawa matrices,
thus the supersymmetric flavor problem is solved by a ``super-GIM''
mechanism~\cite{hallrandall}.

Gaugino domination also alleviates the supersymmetric CP problem
because the only sources for new phases are the gaugino masses,
$\mu$, and $B$. Two phases can be rotated away by field re-definitions,
leaving only one possible new phase. For the special case $B=0$
the supersymmetric CP problem is solved automatically. 

In fact, the special boundary conditions with vanishing superpartner masses
and A-terms are theoretically well motivated. They arise, for example, 
in no-scale supergravity~\cite{no-scale} or in the recently constructed
higher-dimensional ``gaugino mediation'' models~\cite{KKS,CLNP,MgM}.
In gaugino mediation the MSSM matter fields are confined to a brane
in higher dimensions. Supersymmetry is assumed to break on a distant
parallel brane. Extra dimensional locality forbids direct
couplings between the two branes and thereby suppresses all soft
masses which involve MSSM matter fields (squark and slepton
masses, A-terms). Gauge fields and gauginos propagate in the bulk
and couple directly to supersymmetry breaking, allowing for the
generation of gaugino masses.

Extra-dimensional locality only forbids scalar masses at energies large
compared to the compactification scale $M_{BC}$. At long distances the
theory is four-dimensional and masses are generated from
renormalization as usual. The compactification scale in gaugino
mediation corresponds to a free parameter. Gauge coupling unification
motivates us to choose $M_{BC}>M_{GUT}$, and an upper limit on $M_{BC}$ is
given by the length scale at which Nature becomes non-local,
presumably the string scale or Planck scale. In no-scale supergravity
the scale $M_{BC}$ at which soft scalar masses vanish is related to
the string scale. In the following we will treat $M_{BC}$ as a free
parameter subject to the constraint $M_{GUT} \le M_{BC} \le M_{Planck}$, and
refer to the boundary condition of vanishing scalar masses and A terms
as gaugino mediation.

Clearly this scenario is very appealing and it is crucial to ask if it is
phenomenologically viable. For the particular choice of
$M_{BC}=M_{GUT}$ one finds that
the stau is the lightest superpartner (LSP) which is problematic because
the calculated relic abundance of stable staus exceeds experimental
limits by many orders of magnitude. This observation is often
conceived as a failure of no-scale models and has motivated construction
of models with new fields at intermediate scales to modify the
renormalization group equations.
In this paper we repeat the analysis for general $M_{BC} \ge M_{GUT}$
and find the good news that the problematic stau-LSP is very special to
$M_{BC} \approx M_{GUT}$.
For compactification scales slightly higher than $M_{GUT}$
the stau mass gets a large new contribution from running
in the unified theory above $M_{GUT}$ which lifts
its mass above the mass of the Bino. This results in a
very satisfying cosmological picture with a Bino-LSP.

Usually, the renormalization above the GUT scale is ignored (see, however,
Refs.~\cite{PP,aboveGUT,BHS,baer}). There are two seemingly good
reasons for this negligence, we find that both
are invalid. The first reason given is that  
$\log(M_{BC}/M_{GUT})$ is negligibly small compared to
$\log (M_{GUT}/M_{weak})$. However, we find that the smallness of the logarithm
is compensated for by much larger group theory factors
which arise in GUT theories. In particular, the right-handed sleptons
receive only very small contributions from running below the 
GUT scale because they carry only hypercharge, but above $M_{GUT}$
they are unified into a much larger representation with large
corresponding group theory factors. 
The second reason is that the renormalization group equations above
$M_{GUT}$ are necessarily model-dependent. 
Obviously, GUTs have more fields than just the MSSM multiplets and the adjoint
field needed to break the GUT symmetry to the Standard Model gauge group.
For example, there must be additional multiplets that guarantee the splitting
of the doublet and triplet Higgs fields. All such new GUT-scale fields appear
with model-dependent SUSY couplings. However, this does not necessarily imply
that the running of all soft parameters above the GUT scale is model dependent.
As we will demonstrate in gaugino mediation most of the soft masses
decouple from the unknown physics and many predictions can be made
in a model-independent fashion.

In the next two sections we describe the renormalization group
analysis above (Section~\ref{aboveGUT}) and below
(Section~\ref{belowGUT}) $M_{GUT}$ in detail.
We find that the number of model-independent
predictions that can be obtained is related to the form of the
boundary conditions at $M_{BC}$.  
General gaugino mediation boundary conditions allow arbitrary soft
Higgs masses~\cite{CLNP,KT}, a B-term and gaugino
masses at the scale $M_{BC}$. We
show that in this case the spectrum of the first two
generations and the ratios of gaugino masses can be predicted.
When only $M_{1/2}$ and $B$ are non-zero at $M_{BC}$ one can also predict
third generation scalar masses.
Finally, when $B$ is also set to zero one obtains the boundary
conditions of Minimal Gaugino Mediation~\cite{MgM}. In this case the
entire superpartner spectrum can be predicted in terms of just two
parameters: $M_{1/2}$ and $M_{BC}$.
We conclude in Section~\ref{discussion}.

\section{Renormalization above the GUT scale}
\label{aboveGUT}

In this section we discuss the renormalization of soft mass
parameters above the grand unification scale, which we take
to be $M_{GUT}=2\ 10^{16}$ GeV. We first briefly
summarize the situation in a general SUSY-GUT with soft masses
and then turn to gaugino dominated scenarios
such as gaugino mediation or no-scale supergravity.
We assume that there is a direct coupling
of supersymmetry breaking to the gauginos. Then the (one-loop)
``anomaly mediation'' \cite{anomalymediation} contributions
to superpartner masses are always negligible. 

\subsection{General case}

A realistic GUT theory
requires a number of new fields above the GUT scale for breaking the
GUT symmetry, splitting the Higgs doublets and triplets, and possibly
also for generating flavor. There exists a large number of different
proposals for addressing all these problems, but unfortunately 
present day experiments do not allow us to single out a
unique ``GUT Standard Model''.
Not knowing the exact spectrum and couplings above the GUT
scale makes it impossible to perform a reliable renormalization
group calculation of all superpartner masses above the GUT scale.
A conservative approach would then be to parameterize
our ignorance by assuming general non-universal but GUT-symmetric
superpartner masses at $M_{GUT}$. To simplify and to avoid
conflicts with experimental bounds on flavor violation
one often assumes that the soft parameters are approximately
flavor symmetric. This approximation can be poor for
third generation scalar masses which can be significantly modified
because of the large Yukawa couplings. However, the first and second
generation Yukawa couplings are presumably small also above $M_{GUT}$
and can therefore be neglected in the running. Then the one-loop
renormalization of the soft scalar masses for the first and second
generation depends only on the unified gauge coupling and is flavor-universal.
It is therefore possible to compute the running of these soft masses
and one can make predictions if they are known at some
high scale. The results of this running are well-known \cite{martin}.

\subsection{Gaugino domination}

In gaugino-dominated scenarios the situation is different. In gaugino
domination one assumes that the scalar masses as well
as the tri-linear soft terms vanish at some high scale $M_{BC}$, and
their low-energy values are generated from the renormalization group.
Thus at the scale $M_{BC} \ge M_{GUT}$ the non-vanishing mass parameters are
the universal gaugino mass $M_{1/2}$ and the supersymmetry preserving
and violating Higgs mass parameters $\mu$ and $B$. In addition, one
could also have soft masses for the Higgs fields, but for
the time being let us assume that $m_{H_u}$ and $m_{H_d}$ are zero at
$M_{BC}$.

As we will see below, $\mu$ does not enter any renormalization group
equation for the soft terms neither above nor below $M_{GUT}$.
Since we do not have a physical
principle which tells us the value of $\mu$ at the high scale, this
means that there is no need to run the value of $\mu$.
In the phenomenological analysis we simply work with its low-energy value.
$B$ also does not enter any renormalization group equations. However,
Minimal Gaugino Mediation predicts $B=0$ at the high scale, it is
therefore important to predict the low-energy value of $B$ from the
renormalization group running between $M_{BC}$ to $M_{weak}$.
 
The evolution of
the unified gaugino mass between $M_{BC}$ and $M_{GUT}$ is easily
determined because at one loop $M_{1/2}$ simply tracks the evolution
of the unified gauge coupling
\beq
\ddt {1\over g^2}=-2b_{GUT} \qquad \ddt {M_{1/2} \over g^2} =0 \, .
\label{gaugrun}
\eeq
Here we defined $t=1/(16\pi^2) \log(M/M_{GUT})$.
The renormalization group equations for all other soft masses above $M_{GUT}$
are\footnote{We define the soft mass parameters $A$ and $B$ such that
they multiply the Yukawa couplings and $\mu$ in the Lagrangian,
respectively. For example, ${\cal L}_{soft} \sim
A_t Y_t \, {\bf 10} \, {\bf 10} \, {\bf 5}_{H_U} +
B \mu \, {\bf 5}_{H_U} {\bf \bar{5}}_{H_D}\, $.} 
\bea
\ddt m^2_{10} &=& \frac{3}{2} \ddt m^2_{\bar 5} = \frac{3}{2} \ddt m^2_5
= - \frac{144}{5} g^2 M^2_{1/2}\, , \\
\ddt A_u &=& \frac{8}{7} \ddt A_d = 2 \ddt B = \frac{192}{5} g^2 M_{1/2}\, ,
\label{rgesu5}
\eea
for the case of $SU(5)$ and
\bea
\ddt m^2_{16} &=& \frac{5}{4} \ddt m^2_{10} = - 45 g^2 M^2_{1/2}\, , \\
\ddt A_{u,d} &=& \frac{7}{4} \ddt B = 63 g^2 M_{1/2} \, ,
\label{rgeso10}
\eea
for $SO(10)$. On the right-hand side of these equations we assumed
that all soft masses except $M_{1/2}$ are negligible. This is a
very good approximation at energies near $M_{BC}$ where they all
vanish but becomes worse if significant scalar masses are
generated from the renormalization group running.

Note that all dependence on unknown couplings above the GUT scale
has dropped out of Eqs.~(\ref{rgesu5}) and (\ref{rgeso10}).
The remaining model
dependence lies in the choice of grand unified group and in
the evolution of the GUT gauge coupling (specifically $b_{GUT}$).

The solutions to these renormalization group equations are most easily
written by using Eq.~(\ref{gaugrun}) to replace $M_{1/2}(\mu)
\rightarrow M_{1/2}(M_{GUT}) \alpha(\mu)/\alpha_{GUT}$ and defining
\bea
I_4 &=& \int_0^{t_{BC}} g^4 dt
     = \alpha^2_{GUT}\ \log({M_{BC}\over M_{GUT}})
{1\over 1-{b_{GUT}\alpha_{GUT}\over 2\pi} \ \log({M_{BC}\over M_{GUT}})}\, , 
\label{eye4} \\
I_6 &=& \int_0^{t_{BC}} g^6 dt 
     = 4 \pi \alpha^3_{GUT}\ \log({M_{BC}\over M_{GUT}})
     {1-{b_{GUT}\alpha_{GUT}\over 4\pi}\ \log({M_{BC}\over M_{GUT}})\over
     (1-{b_{GUT}\alpha_{GUT}\over 2\pi}\ \log({M_{BC}\over M_{GUT}}))^2} \, .
\label{eye6}
\eea
We find for $SU(5)$
\bea
m^2_{10} &=& \frac32 m^2_{\bar 5} = \frac32 m^2_5
= \frac{144}{5} M^2_{1/2}\ I_6\, , \\
A_u &=& \frac87 A_d = 2 B = - \frac{192}{5} M_{1/2}\ I_4 \, ,
\eea
and for $SO(10)$
\bea
m^2_{16} &=& \frac{5}{4} m^2_{10} = 45\ M_{1/2}^2\ I_6 \, , \\
A_{u,d} &=& \frac{7}{4} B = - 63\ M_{1/2}\ I_4 \, ,
\eea
where from now on $M_{1/2}$ stands for the unified gaugino mass evaluated
at $M_{GUT}$. 

Now all soft supersymmetry breaking at the GUT scale is
determined in terms of the five parameters $M_{1/2}$, $\mu$,
$B$, $I_4$ and $I_6$. $I_4$ and $I_6$ are given in terms
of the two parameters $b_{GUT}$ and $\log(M_{BC}/M_{GUT})$. 
Note that $I_4$ and $I_6$ are small enough for our 
approximation (ignoring all soft masses except for the gaugino mass) to
be valid unless the denominators in Eqs.(~\ref{eye4}) and (\ref{eye6})
go to zero.
But the same vanishing denominators also appear in the evolution of
the gauge coupling. Therefore our approximation is good if and only
if the theory stays perturbative up to the mass scale $M_{BC}$.  
For example, for ${M_{BC}\over M_{GUT}}=10$ perturbativity allows
a beta function coefficient $b_{GUT}$ as large as 50. Note that
this leaves sufficient room for non-minimal Higgs sectors above
$M_{GUT}$ because the minimal $SU(5)$ theory only has $b_{GUT}=-3$,
an extra generation adds $2$ and extra adjoint superfields
contribute 5.
If the theory above the GUT scale has large new Yukawa couplings,
for example a coupling of $H_u$ and $H_d$ to a GUT adjoint superfield, then
terms proportional to scalar masses on the right hand side can make
contributions to scalar masses which become important for large $M_{BC} \sim
M_{Planck}$. This can be seen from the numerical solutions
to the renormalization group equations of the minimal $SU(5)$ model with
gaugino mediation boundary condition presented in Ref.~\cite{baer}.

This discussion changes if we allow
non-zero supersymmetry violating masses $m^2_{H_u}$ and $m^2_{H_d}$
at $M_{BC}$. Not only is the running of these masses very
sensitive to new physics above the GUT scale (such as the
doublet-triplet splitting mechanism or a non-minimal GUT Higgs
sector), but also the third generation scalar masses are now
model dependent because of contributions proportional to the large Yukawa
couplings. Thus, in this case model-independent predictions are only
possible for the first and second generation scalar masses.

\section{The spectrum}
\label{belowGUT}

In this section we discuss the superpartner masses which
result from running from the compactification scale all the way down
to the weak scale. Throughout the discussion we assume that the
gravitino mass is larger than the gaugino masses. This assumption
is important for the phenomenology of the model. Parametrically, in
gaugino mediation one finds $m_{3/2}=\sqrt{V} m_{1/2}$
where $V>8$ is the volume of the extra dimensions in fundamental
Planck units \cite{KKS,CLNP}.
We proceed by presenting three qualitatively
different scenarios. All three scenarios have vanishing squark and
slepton masses and A-terms at the high scale but they differ in the assumptions
made about the Higgs sector. In scenario {\bf A} we allow general soft
Higgs mass parameters $m_{H_u}^2$, $m_{H_d}^2$,
$B$, and $\mu$. In {\bf B}
we specialize to models with vanishing non-holomorphic masses
$m_{H_u}^2 = m_{H_d}^2= 0$. And in {\bf C} we also assume $B=0$,
so that the only remaining mass parameters of the model are $M_{1/2},
\mu, M_{BC}$ and $M_{GUT}$. In the last section we also give three
example superpartner spectra corresponding to representative sets
of model input parameters which satisfy the \MgM\ boundary conditions.

We solve the one-loop renormalization group equations of the MSSM~\cite{MV}
below the GUT scale numerically. We include the effects of third
generation Yukawa couplings and neglect the smaller Yukawa couplings.
The GUT scale boundary conditions for all couplings follow from the
analysis in Section II. Note that there are no threshold
contributions to the supersymmetry breaking parameters in the $\overline{DR}$
renormalization scheme from integrating out the heavy GUT
gauge bosons and gauginos. Explicitly, this follows because diagrams
renormalizing the scalar masses with heavy GUT gauginos in a loop
have vanishing finite pieces in $\overline{DR}$.
At the weak scale we use a one-loop improvement for the Higgs
potential~\cite{HempflingHaber,PBMZ} which captures the effect of
top loops below the stop mass threshold. The top loops
modify the coefficient of the $( H_u^\dagger H_u)^2$ quartic term
and represent the dominant
correction to the mass of the lightest Higgs particle.
The accuracy of this approximation is to better than 10
GeV~\cite{HHH}, when the
running top quark mass $m_{top}(m_{top})$ is used for the calculation.
 
\subsection{General Higgs mass parameters}

When we allow Higgs masses $m_{H_u}^2$ and $m_{H_d}^2$ 
at the scale $M_{BC}$ then only the soft masses of the
first two generations and gaugino masses
can be predicted without knowledge of
details of the GUT physics. 
It is convenient to use the average
soft Higgs mass $(m_{H_u}^2+m_{H_d}^2)/2$ and the difference
$m_{H_u}^2-m_{H_d}^2$ as input parameters. The average Higgs
mass does not contribute directly to first and second generation
scalar masses. Indirectly, it does contribute to scalar masses
through weak-scale D-terms, but for large enough $\tan \beta$
these D-terms can be written universally in terms of the W and Z masses. 
The difference does contribute because it generates a
D-term for hypercharge. This D-term is proportional to the
renormalization group invariant quantity $S$ defined as
\beq
S=m_{H_u}^2-m_{H_d}^2+{\rm Tr}(m_Q^2-2m_U^2+m_E^2+m_D^2-m_L^2)
 =\left. m_{H_u}^2-m_{H_d}^2\, \right|_{M_{GUT}}\ ,
\eeq
where the second equality is valid only at the GUT scale because
the squark and slepton masses in the trace are GUT symmetric and
therefore drop out of the equation. The D-term mass shift for
each scalar at the weak scale is simply
\beq
\delta m_i^2= - \frac65\, y_i\,  S \int_{t_{weak}}^0 g_1^2 \, dt
         \simeq - .078\, y_i\, S \ .
\label{dterm}
\eeq

\begin{figure}[!ht]
\PSbox{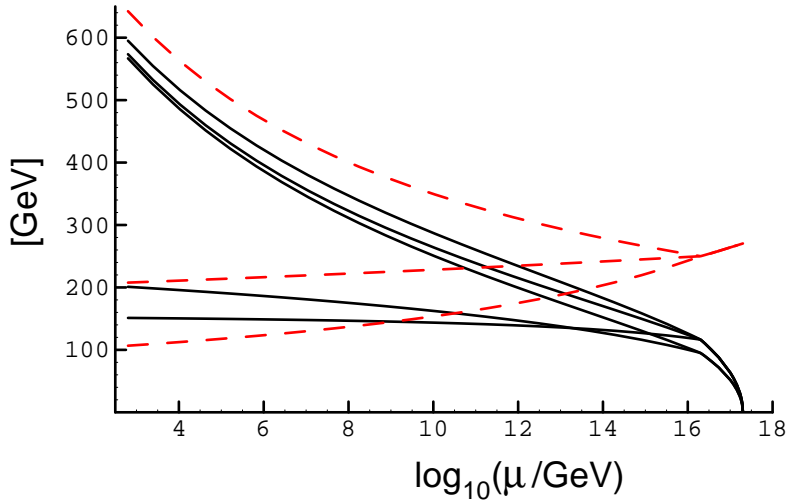 hscale=95 vscale=95 hoffset=65  voffset=0}{13.7cm}{6.5cm}
\caption{Evolution of the soft masses of the first two generations (solid)
and the gauginos (dashed) as a function of renormalization scale
$\mu$. The input parameters are $M_{1/2}=250$~GeV,
$M_{BC}=2\cdot 10^{17}$~GeV, and vanishing hypercharge D-term $S=0$.
The scalar fields are, from the lightest to
heaviest at the weak scale, the right-handed selectron,
the left-handed sleptons, the right-handed down and up squarks,
and the left-handed squarks. The gaugino masses start at a nonzero value
at $M_{BC}$. At the weak scale the gluino is heaviest and the
Bino lightest.}
\label{fig:allrunning}
\end{figure}

We first specialize to the case with no D-term for hypercharge, i.e.
$m_{H_u}^2=m_{H_d}^2$ at $M_{GUT}$. 
Figure~\ref{fig:allrunning} illustrates the evolution of soft masses for
the first two generations and the gauginos from $M_{BC} > M_{GUT}$ to
the weak scale. At $M_{BC}$ the soft
masses vanish and evolve according to $SU(5)$ RGEs down to $M_{GUT}$.
For the purpose of this plot we assumed $SU(5)$ unification
therefore the $\bar{\bf{5}}$ and $\bf{10}$ evolve at different rates.
The gaugino masses are unified
between $M_{BC}$ and $M_{GUT}$. Below $M_{GUT}$ the RGEs respect only the
symmetries of the Standard Model. The evolution depends on the gauge charges
of the fields. Gaugino masses and scalar masses are proportional
to the squares of the gauge couplings.
As a result, colored fields are always heaviest and have masses
about four times larger than fields with only hypercharge.

\begin{figure}[!ht]
\PSbox{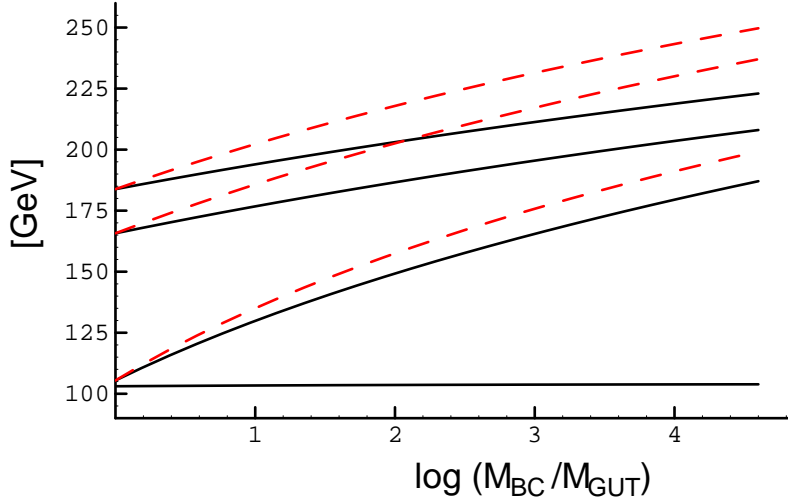 hscale=95 vscale=95 hoffset=65  voffset=0}{13.7cm}{6.5cm}
\caption{Weak scale superpartner masses as a function of
$\log(M_{BC}/M_{GUT})$.
The common gaugino mass is 250~GeV at $M_{GUT}$ and we take $S=0$.
The lightest particle is the neutralino, its mass is independent
of $M_{BC}$. The sleptons are, from the lightest to the heaviest,
the right-handed selectron, the left-handed sneutrino, and the left-handed
selectron. The solid lines correspond to the running in $SU(5)$,
dashed lines correspond to $SO(10)$.}
\label{fig:sleptons}
\end{figure}

The effects of the running above the GUT scale are depicted in
Figure~\ref{fig:sleptons}. Scalar masses receive additive GUT-symmetric
contributions from the running above
$M_{GUT}$. This effect is most important for scalars which do
not receive large masses from running below the GUT scale.
The mass shifts are proportional to the Casimirs of the
corresponding unified representations, thus they are larger in $SO(10)$
compared to $SU(5)$.

\begin{figure}[!hb]
\PSbox{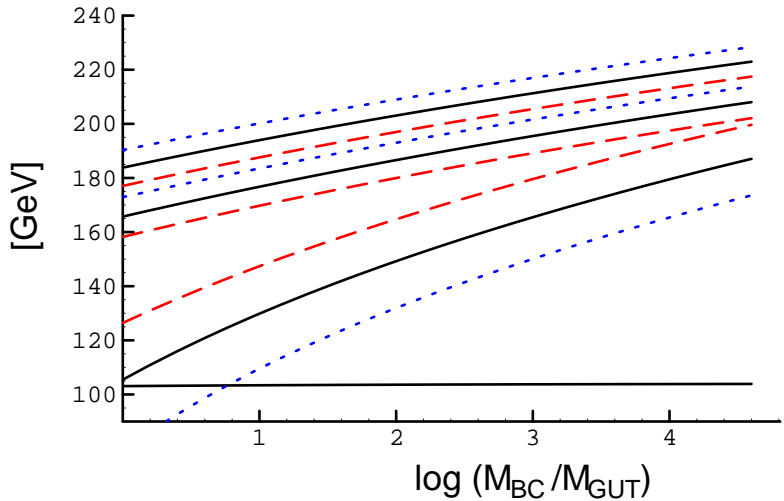 hscale=95 vscale=95 hoffset=65  voffset=0}{13.7cm}{6.5cm}
\caption{Weak scale superpartner masses as a function of
$\log(M_{BC}/M_{GUT})$ for three different values of
$S=-M_{1/2}^2, 0, +M_{1/2}^2$. 
We take $M_{1/2}=250$ GeV and GUT group $SU(5)$.
The lightest particle is the neutralino, its mass is independent
of $M_{BC}$. The other solid lines correspond to, from lightest to
heaviest, the right-handed selectron, the left-handed sneutrino,
and the left-handed selectron for $S=0$, as in
Figure~\ref{fig:sleptons}. Dashed lines correspond to $S=-M_{1/2}^2$
and dotted lines to $S=+M_{1/2}^2$. Note that left- and right-handed
slepton masses are shifted in opposite directions.}
\label{fig:dterm}
\end{figure}

A non-vanishing D-term introduces the additional input parameter $S$.
The $S$-dependence of the first and second generation scalar masses is easily
accounted for by using equation~(\ref{dterm}). The effects of the D-term
are largest for the lightest superpartners. In Figure~\ref{fig:dterm}
we show the slepton masses for the case of $SU(5)$ GUT group and
three different values of $S$. The mass shifts from the
hypercharge D-term are in opposite directions for left- and right-handed
sleptons.

Note that the hypercharge D-terms also contribute to stau masses
and may be responsible for lifting the right-handed stau mass above
the Bino mass in models where $M_{BC}=M_{GUT}$. This has been
used in Refs.~\cite{CLNP,KT}.
 
\subsection{$m_{H_u}^2 = m_{H_d}^2= 0$}

When the Higgs masses are zero at $M_{BC}$, the Yukawa couplings do
not significantly contribute to the running above the GUT scale.
Therefore the running for all three generations can be computed
model independently. For fixed $M_{1/2}$ and $M_{BC}$,
$B$ and $\tan \beta$ are related at the minimum of
the Higgs potential. Thus, we can express $B$ in terms of $\tan \beta$
or vice versa.

\begin{figure}[!ht]
\PSbox{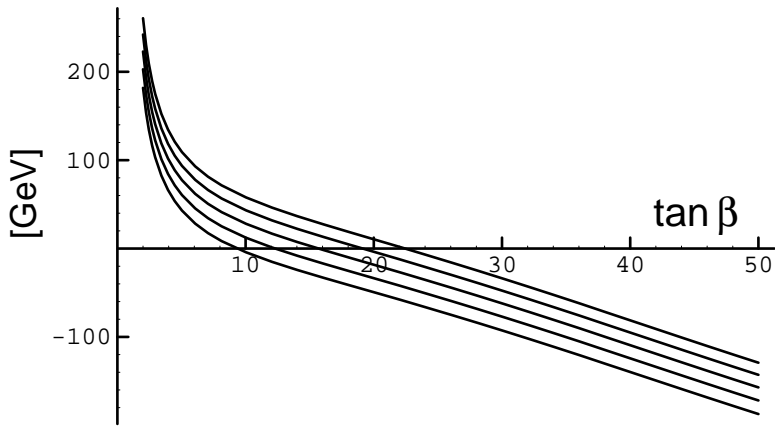 hscale=95 vscale=95 hoffset=65  voffset=0}{13.7cm}{6.5cm}
\caption{The dependence of $B$, evaluated at $M_{BC}$, on $tan \beta$.
We take $M_{1/2}=250$ GeV, the different curves correspond
to $\log(M_{BC}/M_{GUT})=0,1,2,3,4$ from left to right, respectively.}
\label{fig:B-tanbeta}
\end{figure}

The relation between $B$ at the high scale and $\tan \beta$
is depicted in Figure~\ref{fig:B-tanbeta} for different values of $M_{BC}$.
It is clear that a value of $B$ can be picked for any values
of $\tan \beta$ and $M_{BC}$. It is therefore more convenient to
treat $\tan \beta$ as an input parameter, as is usually done in
analyzing supersymmetric theories.

\begin{figure}[!ht]
\PSbox{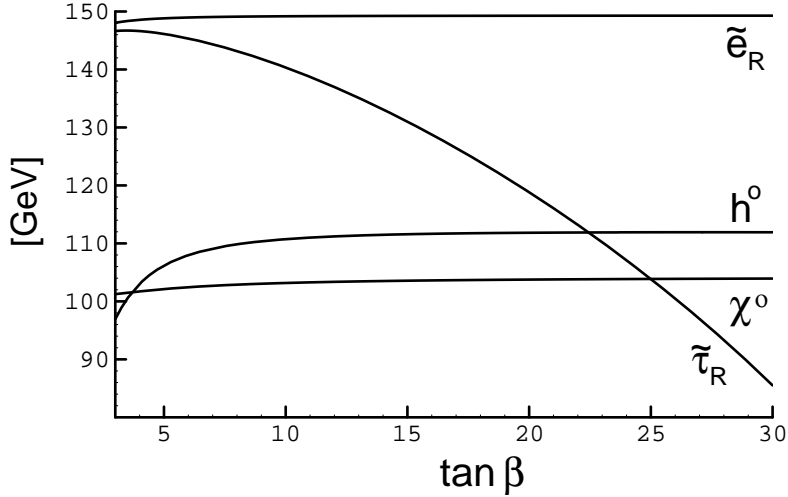 hscale=95 vscale=95 hoffset=65  voffset=0}{13.7cm}{6.5cm}
\caption{Masses of the right-handed selectron, stau, the lightest Higgs
and neutralino as a function of tan beta. The common gaugino mass
is 250~GeV at $M_{GUT}$ and $\log(M_{BC}/M_{GUT})=2$.}
\label{fig:sleptons3}
\end{figure}

Since the ratio of the bottom and top Yukawa couplings depends on
$\tan \beta$ the masses of the third generation particles vary
with $\tan \beta$. In particular, the mixing between left- and
right-handed sleptons increases with $\tan \beta$.  As a result
one of the mass eigenstates becomes lighter with increasing $\tan \beta$.
Figure~\ref{fig:sleptons3} illustrates this strong dependence in
case of the right-handed stau.

\subsection{$m_{H_u}^2 = m_{H_d}^2= 0$, $B=0$}

In minimal gaugino mediation all soft terms except
for $M_{1/2}$ (and $\mu$) vanish at $M_{BC}$. The values of
$\tan \beta$ which correspond to $B=0$ can be determined easily from
Figure~\ref{fig:B-tanbeta} as a function of $M_{BC}$.
The two remaining parameters are $M_{1/2}$ and the
scale $M_{BC}$. With only two free parameters the theory is highly predictive.
To a good approximation $M_{1/2}$ sets the linear scale of all superpartner
masses.\footnote{Note that the lightest Higgs mass does not scale linearly
with $M_{1/2}$, because at tree level the Higgs mass is bounded
by $M_Z$. Radiative corrections to the Higgs potential from top
and stop loops introduce a logarithmic dependence on $M_{1/2}$.}
The second parameter, $\log (M_{BC} / M_{GUT})$, increases
the slepton masses relative to gaugino masses. A more detailed discussion of
\MgM\ is contained in Refs.~\cite{MgM,baer}.

For demonstration, we present the spectra for three sample points
of the parameter space in Table~\ref{studytable}.
The first point with $M_{1/2}=200$~GeV (``light
\MgM'' scenario) is in the lower range of experimentally
allowed values of $M_{1/2}$.
This point will be probed in the near future by the ongoing
Higgs search at LEP II. The lighter chargino and the second lightest
neutralino can be observed at the Tevatron in the
$pp\rightarrow \chi^\pm_1 \chi^0_2 \rightarrow 3 l$ channel.
While the chargino and neutralino in the ``light'' scenario might evade
Run II they are certainly within the reach of Run III~\cite{Tevatronworking}.
Our second and third points,
with $M_{1/2}=300, 500$~GeV -- ``intermediate \MgM ''
and ``heavy \MgM '' scenarios -- cannot be
tested at LEP II, but the lightest Higgs is within the
reach of Run II at the Tevatron.  
The superpartners corresponding to the second and third 
study points are too heavy to be seen at LEP or the
Tevatron, but they are easily within the reach of both LHC and NLC.

One can set an upper bound on $M_{1/2} \lessim 600$ GeV by requiring that
the relic abundance of \MgM\ Bino-LSPs does not contribute more than
0.1 to 0.3 to critical density~\cite{MgM}.
We find that in our ``light'' scenario Binos are not abundant enough to
account for all cold dark matter, whereas the ``intermediate'' and
``heavy'' scenarios are in the preferred region.

\begin{table}[t]
\begin{center}
\begin{tabular}{cccc|cccc|cccc}
\multicolumn{4}{c}{I. ``light \MgM''}&
\multicolumn{4}{c}{II. ``intermediate \MgM''}&
\multicolumn{4}{c}{III. ``heavy \MgM''}\\
\hline \hline
Field  & mass \qquad &Field & mass &
Field  & mass \qquad &Field & mass &
Field  & mass \qquad &Field & mass \\ \hline
$\wtil g$               & 520   &                       &       &
$\wtil g$               & 780   &                       &       &
$\wtil g$  	        & 1300  &                       &       \\
$\wtil \chi_1^\pm$      & 147   & $\wtil \chi_2^\pm$    & 317   &
$\wtil \chi_1^\pm$      & 235   & $\wtil \chi_2^\pm$    & 457   &
$\wtil \chi_1^\pm$	& 406   & $\wtil \chi_2^\pm$    & 750   \\
$\wtil \chi_1^0$        & 82  	& $\wtil \chi_2^0$      & 148   &
$\wtil \chi_1^0$        & 125  	& $\wtil \chi_2^0$      & 236   &
$\wtil \chi_1^0$	& 211 	& $\wtil \chi_2^0$      & 406   \\
$\wtil \chi_3^0$        & 294   & $\wtil \chi_4^0$      & 315   &
$\wtil \chi_3^0$        & 441   & $\wtil \chi_4^0$      & 456   &
$\wtil \chi_3^0$   	& 739   & $\wtil \chi_4^0$      & 750   \\
$\wtil u_L$             & 478   & $\wtil u_R$           & 463   &
$\wtil u_L$             & 720   & $\wtil u_R$           & 696   &
$\wtil u_L$	        & 1188  & $\wtil u_R$           & 1145  \\
$\wtil d_L$             & 485   & $\wtil d_R$           & 460   &
$\wtil d_L$             & 725   & $\wtil d_R$           & 689   &
$\wtil d_L$		& 1191  & $\wtil d_R$           & 1138  \\
$\wtil t_1$             & 351   & $\wtil t_2$           & 501   &
$\wtil t_1$             & 535   & $\wtil t_2$           & 703   &
$\wtil t_1$	        & 906   & $\wtil t_2$           & 1117  \\
$\wtil b_1$             & 437   & $\wtil b_2$           & 497   &
$\wtil b_1$             & 664   & $\wtil b_2$           & 735   &
$\wtil b_1$	        & 1121  & $\wtil b_2$           & 1195  \\
$\wtil e_L$             & 167   & $\wtil e_R$    	& 126   &
$\wtil e_L$             & 245   & $\wtil e_R$    	& 183   &
$\wtil e_L$	        & 373   & $\wtil e_R$           & 235   \\
$\wtil \nu_e$           & 147   & $\wtil \nu_\tau$      & 147   &
$\wtil \nu_e$           & 231   & $\wtil \nu_\tau$      & 231   &
$\wtil \nu_e$	        & 364   & $\wtil \nu_\tau$      & 364   \\
$\wtil \tau_1$		& 104   & $\wtil \tau_2$        & 178   &
$\wtil \tau_1$		& 162   & $\wtil \tau_2$        & 253   &
$\wtil \tau_1$    	& 224   & $\wtil \tau_2$        & 375   \\
$h^0$                   & 108   & $H^0$       		& 294   &
$h^0$                   & 115   & $H^0$       		& 453   &
$h^0$		        & 122   & $H^0$                 & 793   \\
$A^0$                   & 294   & $H^\pm$               & 312   &
$A^0$                   & 453   & $H^\pm$               & 464   &
$A^0$		        & 793   & $H^\pm$               & 800   \\
\hline \hline
\end{tabular}
\end{center}
\caption{\em Masses of superpartners, in GeV, for \MgM\ study points
I, II and III. They correspond to parameter values 
``light'' ($M_{1/2}=200$ GeV, $M_{BC}/M_{GUT}=10$, $\tan\beta=17$),
``intermediate'' ($M_{1/2}=300$ GeV, $M_{BC}/M_{GUT}=10$, $\tan\beta=17$),
and ``heavy'' ($M_{1/2}=500$ GeV, $M_{BC}/M_{GUT}=2$, $\tan\beta=12$),
respectively.
\label{studytable}}
\end{table}

\section{Conclusions}
\label{discussion}

We conclude that the MSSM with gaugino mediated supersymmetry breaking
is not only phenomenologically viable, but it also has a number of very
attractive features:
\begin{itemize}

\item The model solves the supersymmetric flavor problem
because the squark and slepton masses which are generated from the
renormalization group evolution are sufficiently degenerate (and
aligned~\cite{NS}).
  
\item It is theoretically well motivated. The vanishing of the
scalar masses and the $A$ terms at high scales is a natural prediction of
models with extra dimensions where MSSM gauge fields are bulk fields,
whereas the MSSM matter fields and the supersymmetry breaking
mechanism are localized on separate branes.

\item The Bino-LSP of these models makes a good cold dark matter
candidate. Cosmic abundances in the range $\Omega h^2 = .1 - .3$
are obtained for right-handed selectron masses in the range $150 - 250$~GeV
as shown in \cite{MgM,cdm}. 

\item The model is very predictive because superpartner masses depend
only on a small number of input parameters. For example, in \MgM\
all superpartner masses can be computed in terms of two parameters
$M_{1/2}$ and $\log (M_{BC}/M_{GUT})$. In the more general case with
non-vanishing soft Higgs masses at $M_{BC}$ all gaugino masses and the
first and second generation scalar masses are predicted in terms of the same
two parameters and possibly a hypercharge D-term $S$.

\end{itemize}
In our analysis of the model the renormalization group running above
the GUT scale was essential for determining the masses of the lightest
scalar superpartners. Its most important effect is that it raises the stau
mass above the mass of the Bino.
Precise measurements of the superpartner masses would allow a determination
of the small contributions from running above the GUT scale
\cite{MW}. Consistency with the gaugino mediation
predictions would constitute a decisive test of the
scenario and allow an indirect measurement of the GUT gauge group.

The framework is predictive because at the one-loop level above-the-GUT-scale
model dependence decouples from the soft supersymmetry breaking terms.
This is a consequence of vanishing
scalar masses at $M_{BC}$ and would not be true for more general SUSY GUTs.
If the soft Higgs mass parameters $m_{H_u}^2$ and $m_{H_d}^2$ are
non-vanishing at the GUT scale then the third generation scalar masses
and Higgs masses become model-dependent, but first and second generation
scalar masses as well as gaugino masses can still be predicted. In this case
the hypercharge D-term proportional to $m_{H_u}^2-m_{H_d}^2$ also
plays an important role in determining the lightest superpartner masses.

We have left studies of the collider phenomenology of these models for
future work. In particular, it would be interesting to determine
the most promising signatures which allow tests of this scenario and the
implications for future colliders. As part of this analysis a more
accurate treatment of the renormalization group (two-loops) and weak
scale threshold effects would allow sharpened predictions for
superpartner masses. This more accurate analysis is necessary for
comparison of our Higgs mass prediction with LEP II bounds. We also expect
interesting predictions for (and constraints from) flavor violating
transitions such as $b \rightarrow s \gamma$ and $\mu \rightarrow e \gamma$.

\acknowledgements

We thank Howard Baer, Markus Luty, Elazzar Kaplan, Konstantin Matchev,
Michael Peskin, Raman Sundrum, and Jim Wells for useful discussions.
MS is supported by the DOE under contract DE-AC03-76SF00515.
WS is supported by the DOE under contract DOE-FG03-97ER40506.

\end{document}